\documentclass[nofootinbib,reprint, amsmath,amssymb,aps, superscriptaddress]{revtex4-2}

\usepackage{graphicx}
\usepackage{dcolumn}
\usepackage{bm}
\usepackage{aas_macros}
\usepackage{amsmath}
\usepackage{comment}
\usepackage[normalem]{ulem}
\usepackage{xcolor}

\begin{document}

\title{Emission and cooling of a newborn quark star}

\author{Mikalai Prakapenia}
\affiliation{Institute of Physics, National Academy of Sciences of Belarus, 220072, Nezalezhnasci Avenue 68-2, Minsk, Belarus}

\author{Gregory Vereshchagin}
\affiliation{ICRANet, 65122 Piazza della Repubblica, 10, Pescara, Italy}
\affiliation{ICRANet-Minsk, Institute of Physics, National Academy of Sciences of Belarus\\
220072 Nezalezhnasci Avenue 68-2, Minsk, Belarus}
\date{\today}

\begin{abstract}
It has long been believed that newborn quark stars are inefficient emitters because the quark plasma frequency is so high that it suppresses electromagnetic radiation across the entire spectrum up to energies of several tens of MeV. Here, we show that significant electromagnetic emission can instead originate from the electrosphere of a quark star, where the plasma frequency is much lower, approximately 40 keV. As a result, the cooling of a newborn quark star is expected to proceed through two distinct observational epochs. During the first epoch, blackbody radiation is emitted from the photosphere of an optically thick relativistic wind generated near the electrosphere. Over the course of a few days following the star’s formation, the photospheric temperature decreases from several MeV to about 22 keV. During the second epoch, the blackbody spectrum evolves into a narrow feature centered around 40 keV, and the emission gradually fades. This spectral evolution provides a characteristic signature of the electrosphere surrounding compact astrophysical objects and may be detectable with current or future X-ray observatories.
\end{abstract}

\maketitle

\textit{Introduction. --} 
The behavior of matter under extreme conditions, such as ultra-high density, pressure, temperature, and field strength, remains one of the central open questions in modern physics. Although laboratory experiments with particle accelerators and ultraintense lasers continue to provide important insights, astrophysical environments remain unique natural laboratories for probing these regimes.

In this context, the strange quark matter (SQM) hypothesis \cite{1984PhRvD..30..272W} is particularly relevant, as it is closely connected to all of these extreme properties. According to this hypothesis, the true ground state of matter consists entirely of deconfined quarks with $u$, $d$, and $s$ flavors present in comparable proportions. A recent study \cite{Holdom2018,2026PhRvD.113d3033Y} suggests that $u,~d$ quark matter (udQM) can be more stable than SQM.

Quark matter may exist in various color-superconducting phases \cite{Alford2008_RMP80-1455}. In two-flavor color-superconducting (2SC) phase $s$ quarks are absent, and $u$ quarks pair with $d$ quarks in antitriplet state leaving one of the colors unpaired. Also, $s$ quarks can appear in an unpaired state forming a 2SC+s phase. In color-flavor locked (CFL) phase, all $u,~d$ and $s$ quarks pair with each other in an antisymmetric pattern. 

A class of hypothetical compact astrophysical objects composed entirely of superdense quark matter is referred to as quark stars \cite{1986ApJ...310..261A,Yuan2022,Restrepo2023}.  
A quark star may form as the core of a massive star undergoes gravitational collapse, producing a supernova \cite{2002A&A...390L..39O,2020ApJ...894....9F,Becerra2026}, or even a gamma-ray burst \cite{2005MNRAS.362L...4P}. Direct observation of a quark star in the early epochs, up to several years after the collapse, is precluded, as the supernova ejecta are opaque to electromagnetic radiation. Alternatively, a quark star may be formed from a neutron star by the phase transition. Such process produces a characteristic neutrino signal \cite{1987PhLB..192...71O,2006PhRvC..74f5804B,2007ApJ...659.1519D,2010PhRvC..82f2801N,2013PhRvD..87j3007P,2015PhRvC..92d5801D,2016PhRvD..93d3018F,2018ApJ...859...57P,zenati2025}, but neutrinos are difficult to detect, and a possible electromagnetic signal is of interest. So we will focus on this astrophysical scenario.


Newborn quark stars are likely to have a bare surface, as the neutrino flux and radiation pressure blow out any hadronic matter from the surface \cite{1992ApJ...391..228W,Usov_2001,2013PhRvD..87j3007P}. Quarks can form  a very sharp surface set up by the strong interactions. Electrons are bound to quarks via electromagnetic interactions and cannot be fully packed inside the quarks. Thus, electrons spread for hundreds of fermis over the quark surface \cite{1986ApJ...310..261A}. Generally speaking, the electron density depends on the QCD phase of the quark matter \cite{Alford2008_RMP80-1455} and the surface tension of the quarks \cite{2005ApJ...620..915U, Xia2017_NPB916-669}.

Charge separation at the surface produces a layer with a supercritical electric field \cite{1986ApJ...310..261A,1995PhRvD..51.1440K} exceeding the Schwinger limit \cite{1951PhRv...82..664S} for pair production $E_c=m_{e}^2c^3/\hbar e \simeq 1.3\times 10^{16}$ V/cm, where $m_{e}$ is the electron mass, $e$ is its charge, $c$ is the speed of light, and $\hbar$ is the reduced Planck constant. This region is called the electrosphere \cite{1998PhRvL..80..230U,2024ApJ...963..149P}.

The quark surface is a poor photon emitter, due to a large plasma frequency (typically about 20 MeV), associated with quarks \cite{1986ApJ...310..261A,2001ApJ...550L.179U,2003ApJ...596..451C}. Above the surface, where quarks are absent, electrons are immersed in a strong electric field. The following elementary processes are possible there: a) Schwinger pair creation in electric field ($E\rightarrow e^+e^-$) \cite{1951PhRv...82..664S,1998PhRvL..80..230U} (referred to in what follows as the \emph{Schwinger mechanism}); b) electron-electron and electron-positron bremmsstrahlung ($ee\rightarrow ee\gamma$) \cite{2004PhRvD..70b3004J,2005ApJ...622.1033H,2009PhRvD..80l3006C}; c) bremmsstrahlung  emission of electrons in \emph{external} (due to charge separation) electric field ($Ee\rightarrow Ee\gamma$) \cite{2010PhLB..690..250Z,2011JETP..112...63Z}. It was shown by Zakharov \cite{2010PhLB..690..250Z,2011JETP..112...63Z} that the process c) (referred to in what follows as \emph{Zakharov mechanism}) actually dominates over the other channels by several orders of magnitude and is comparable to the black body emission at a given temperature. Thus, previous claims about insignificance of photon emission from the surface of a quark star, e.g. \cite{1986ApJ...310..261A,1998PhRvL..80..230U,2005ApJ...620..915U} based mainly on the argument of a large plasma frequency of quarks, appear to be premature. We will therefore neglect other sources of photons besides the Zakharov mechanism. 


In this Letter, we revisit the process of cooling of a newborn quark star taking into account the most relevant mechanisms: neutrino emission from its interior, as well as photon emission by bremsstrahlung in electric field and the Schwinger process of electron-positron pair creation at its surface. 

\textit{Electron-positron pair creation and photon emission in the electrosphere. -- } 
The structure of the electrosphere is determined from the solution of Poisson equation. We assume that the degenerate electron gas is described by the Fermi-Dirac distribution function $f_e$ with given spatially constant temperature $T$ and spatially varying chemical potential $\mu_e$. Then the condition of electrochemical equilibrium $\mu_e=e\varphi$, where $\varphi$ is an electrostatic potential, determines the electron density distribution $n_e$ near the quark star surface. The Poisson equation can be written as $d^2\varphi/d z^2 =  4\pi \alpha  n_e(\varphi)$, where $z$ is the spatial coordinate normal to the surface and $\alpha$ is the fine structure constant. To determine the boundary conditions one uses the zero pressure condition $P=0$ for a given equation of state, which gives the quark chemical potential at the surface $\mu_q$. Then the boundary conditions for electrostatic potential are $e\varphi \rightarrow 0$ as $z\rightarrow +\infty$, and $e\varphi \rightarrow\mu_e(P=0)$ as $z\rightarrow -\infty$. We also include surface effects of $s$-quark depletion, which results in a jump of the electric field at $z=0$. This effect depends on the strange quark mass $m_s$ and the quark chemical potential $\mu_q$ \cite{2005ApJ...620..915U, Xia2017_NPB916-669}.

Knowing the distribution of electric field and chemical potential of electrons, one can compute the luminosity of pairs created via the Schwinger mechanism. We adopt the
differential pair creation rate given by Gatoff et al. \cite{1987PhRvD..36..114G} with an additional Pauli blocking factor $(1-f_e)$. The rate is given by an integral over particle phase space
\begin{gather}
\label{ndotschwinger}
\frac{dn_\pm}{dt} = - \frac{E/E_c}{2\pi^2} \int\limits_0^\infty d p_\perp~ p_\perp \left(1-f_e\right) \\ \notag \times\ln\left[  1-\exp\left(-\frac{\pi( p_\perp^2+m_e^2)}{E/E_c} \right)\right],
\end{gather}
where $p_\perp$ is electron momentum orthogonal to the electric field, i.e. $z$ direction.
Then the Schwinger luminosity is obtained \cite{prakapenia2026} as
\begin{gather}
L_\text{Schwinger} = 4\pi R^2 \gamma m_e\int\limits_0^{z_0} \frac{dn_\pm}{dt} dz, 
\label{Lschwinger}
\end{gather}
where $\gamma$ is the Lorentz factor of positrons, $R$ is the radius of the star, and $z_0$ is the distance from the surface where $\mu_e=m_e$. 

The mechanisms of photon emission in the electrosphere of a quark star are discussed in a number of papers, e.g. \cite{2002ApJ...570L..65X,2004PhRvD..70b3004J,2005ApJ...622.1033H,2009PhRvD..80l3006C}. Zakharov \cite{2010PhLB..690..250Z,2011JETP..112...63Z} was the first to point out the dominance of photon emission by electrons in \emph{external} electric field of the electrosphere over all other energy loss mechanisms, including the Schwinger mechanism for temperatures $0.1$ MeV $<T<1$ MeV. In particular, he demonstrated that the spectrum and luminosity of such photons are comparable to the black body limit.

The photon rate in the Zakharov mechanism is \cite{2011JETP..112...63Z}
\begin{gather}
\frac{dn_\gamma}{dt dk} = \int d^3p f_e(\epsilon)[1-f_e(\epsilon-\omega)] \\ \notag
\times \left( \frac{a}{p\chi}\text{Ai}'(\chi)+\frac{b}{p}\int\limits_\chi^\infty dy \text{Ai}(y) \right),
\end{gather}
where $k$ is photon momentum, $\omega$ is photon energy, $p$ is electron momentum, $\epsilon$ is electron energy and
\begin{gather}
a=-\frac{2N g_1}{M},~~~b=M g_2 - \frac{N g_1}{M},~~~\chi =\frac{N p^{2/3}}{(M x eE p_\perp )^{2/3}},  \notag  \\
g_1 =\alpha x^{-1}(1-x+x^2/2),~~g_2 = \alpha 2^{-1} x^3 m_e^2 M^{-2},  \notag \\
M= p x (1-x),~~N = m_e^2 x^2 +(1-x)m_\gamma^2,~~x=k/p. \notag
\end{gather}
For the photon quasiparticle mass $m_\gamma$ we use the static expression which is the plasma frequency of electrons $m_\gamma=\omega_{pl} = m_D/\sqrt{3}$ with the Debye mass
\begin{gather}
m_D^2 = \frac{4\alpha}{\pi} \left( \mu_e^2 + \frac{\pi^2 T^2}{3} \right).
\end{gather}

The corresponding photon luminosity is
\begin{gather}
\label{ndotzakharov}
L_\text{Zakharov} = 4\pi R^2\int\limits_{0}^{z_0} dz \int\limits_{0}^{\infty} dk \omega(k) \frac{dn_\gamma}{dtdk}.
\end{gather}
It is important to note that the electrosphere is transparent for photons, so photons do not change the structure of the electrosphere determined in \cite{2024ApJ...963..149P}. 
Moreover, a hierarchy of particle and energy densities is established. Electrons are the most abundant particles in the electrosphere. Photons produced by the Zakarov mechanism have a much smaller energy density. Positrons produced via the Schwinger process have even a smaller energy density. When additional electron-positron pairs are produced from photons, e.g. via the Breit-Wheeler process, the number density of these charged particles is still significantly smaller than the number density of electrons in the electrosphere. Therefore, the structure of the electrosphere remains essentially undisturbed by bremsstrahlung emission and electron-positron creation.

\begin{figure}[t]
\includegraphics[width=\columnwidth]{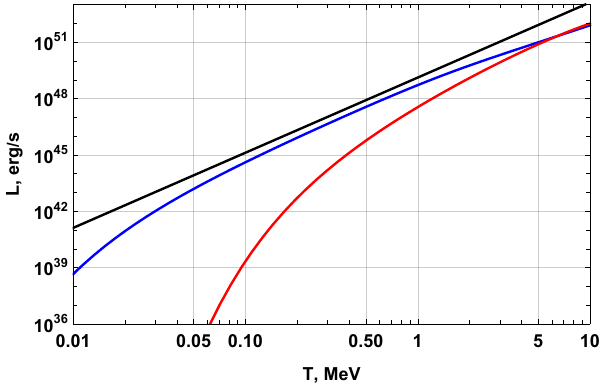}
\caption{Zakharov luminosity (blue), Schwinger luminosity (red) and black body luminosity (black) as a function of temperature $T$.}
\label{Lfitplot}
\end{figure}

The energy loss at the surface of a newborn quark star is due to two processes: Schwinger electron-positron pair creation and Zakharov mechanism of photon generation. The corresponding luminosities are shown in Figure \ref{Lfitplot} as a function of surface temperature, and compared with the black body luminosity. One can see that at high temperatures, $T>5$ MeV, the Schwinger luminosity is comparable to the Zakharov luminosity. Both mechanisms depend on the number of available states for highly degenerate electrons, with the latter decreasing exponentially for low temperatures \cite{1998PhRvL..80..230U}. Therefore, the luminosities of both processes fall exponentially with diminishing temperature. Schwinger luminosity decreases faster, since pairs are produced in highly occupied states with zero component of momentum, parallel to electric field. In contrast, Zakharov luminosity remains comparable to the black body in the range $0.1$ MeV $<T<1$ MeV, and decreases exponentially for lower temperatures.

\begin{figure}[t]
\includegraphics[width=\columnwidth]{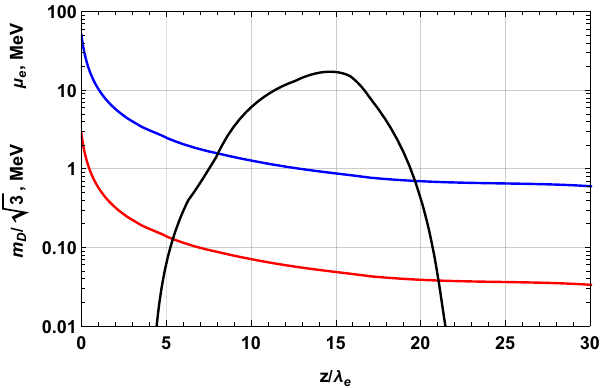}
\caption{Electron chemical potential (blue) and Debye mass (red) in the electrosphere as a function of distance $z$ from the surface of 2SC+s quark star for $T = 0.01$ MeV, $\mu_q =300$ MeV and $m_s=100$ MeV. Black curve represents Zakharov photon rate in arbitrary units. Here $\lambda_e$ is the electron Compton wavelength.}
\label{Zpeak}
\end{figure}

Note that Zakharov \cite{2011JETP..112...63Z} used a method similar to the Arnold-Moore-Yaffe \cite{2003JHEP...01..030A} approach to the collinear photon emission from quark-gluon plasma. This approximation is valid when the photon energies exceed several units of the photon quasiparticle mass (Debye mass), which was adopted as 0.1 MeV by Zakharov \cite{2011JETP..112...63Z}. In Figure \ref{Zpeak} we show the rate $dn_\gamma(z)/dt$ for the Zakharov mechanism as a function of the spatial distance from the surface $z$. The position of the peak of this function depends on the temperature and is shifting away from the surface for a decreasing temperature. The decrease of this function to the left of the peak is due to Pauli blocking of electron states. The decrease in the function to the right of the peak is due to the exponentially decreasing electron density. We note that even for a temperature as low as 0.01 MeV the photons are emitted by relativistic electrons with chemical potential $\mu_e\approx 1$ MeV, which is much larger than the Debye mass (associated with the plasma frequency), so the method of Zakharov is still valid.

\begin{figure}[t]
\includegraphics[width=\columnwidth]{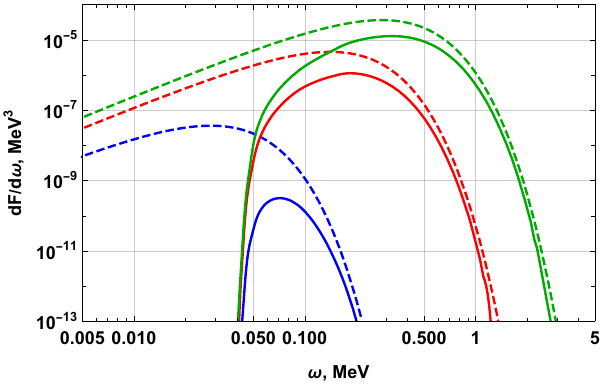}
\caption{Photon emission spectra of the electrosphere for selected temperatures:  $T=0.01$ MeV (blue), $T=0.05$ MeV (red), $T=0.1$ MeV (green). Dashed curves represent the corresponding black body spectra.}
\label{Zspectra}
\end{figure}
The spectrum generated by the Zakharov mechanism is similar to the black body with sharp cut-off at low energies due to the plasma frequency, see Figure \ref{Zspectra}. This low energy cut-off at $\omega_{\min} \approx 40$ keV does not depend on temperature, as the plasma frequency at low energies depends only on the chemical potential of the electrons. For this reason, at sufficiently low temperatures $T< \omega_{\min}$ the spectrum becomes very narrow, centered at $\omega_{\min}$. This feature is mostly determined by the surface effects (see also Fig. 1 from \cite{prakapenia2026a}). We checked that electron chemical potential on the surface $\mu_e(z=0)$ ranges approximately from 50 MeV to 60 MeV for $100~\text{MeV} < m_s < 150~\text{MeV}$ and $300~\text{MeV}<\mu_s<350~\text{MeV}$. Thus, we conclude that the energy $\omega_\text{min}$ is not sensitive to the variations of $m_s$ and $\mu_q$. In addition to the previous paragraph, we note that the peak of the photon spectra with $T=0.01$ MeV lies around $\omega \approx 80$ keV $=2\omega_\text{min}$. Thus, one can conclude that the dominating photon energies exceed two units of the photon quasiparticle mass and the regime of collinear photon emission is adequate.

\textit{Observational appearance of the cooling quark star. --} The cooling of a newborn quark star has been considered in several publications \cite{2002PhRvL..89m1101P,2006NuPhA.774..815B,Fonseca2026}. In general, quark stars cool faster than neutron stars due to intense neutrino emission through direct Urca processes in degenerate quark matter \cite{IWAMOTO19821}.
Recently we revisited the cooling of quark stars \cite{prakapenia2026,prakapenia2026a} paying special attention to the role of the electrosphere producing electron-positron pairs and contributing to cooling via thermal conductivity of quarks. It was shown that a strong temperature gradient is established close to the surface, leading to a fast decrease in surface luminosity down to $10^{40}$ erg/s at 1 s after the birth of a quark star.

In this Letter we consider, as a canonical example, a $1.4M_\odot$ quark star with radius $R=10$ km utilizing three different equations of state of color superconducting quark matter: 2SC, 2SC+s and CFL. The phenomenological equation of state based on the MIT bag model is adopted, which includes four parameters: effective bag constant $B$, gap parameter $\Delta$, strange quark mass $m_s$ and QCD coupling constant $\alpha_s$ \cite{Zhang2021_PRD103-063018}. We construct a stellar configuration as a solution of the Tolman–Oppenheimer–Volkoff equation with the quark chemical potential at the surface $\mu_q \simeq 300$ MeV. To obtain this configuration we fix the parameters as $m_s=100$ MeV, $\Delta=50$ MeV, $\alpha_s=0.1$ and adjust the bag constant having the value about $B^{1/4}\simeq 150$ MeV for all three QCD phases. 

We numerically solve the heat transport equation in spherically-symmetric case $c_v \partial T/\partial t=-r^{-2}\partial(r^2 F_r)/\partial r - Q_\nu$ with a given heat capacity $c_v$, thermal conductivity $\kappa$ and neutrino emissivity $Q_\nu$. For the heat capacity of 2SC and 2SC+s phases, we use the standard expression of relativistic degenerate fermion gas for ungapped quarks, while for CFL phase we have an expression of superfluid phonons, i.e. relativistic massless bosons. We calculate the thermal conductivity of 2SC and 2SC+s phases following the method described in \cite{PhysRevC.90.055205,Alford_2010,Shternin2022}, for CFL phase we use the expression from \cite{braby2010}. Neutrino emissivity in 2SC and 2SC+s phases is dominated by the direct URCA processes of unpaired blue $u$ and $d$ quarks. In CFL phase neutrino emissivity is contributed by two processes: exponentially suppressed quark bremsstrahlung and neutrino emission from Goldstone modes \cite{IWAMOTO19821,JAIKUMAR2001345,Jaikumar2002}. In \cite{prakapenia2026} we argued that neutrino trapping inside the star does not change the cooling behavior of the surface. Therefore, focusing on thermal evolution of the surface we study neutrino-transparent regime utilizing neutrino emissivity $Q_\nu$ in the heat transport equation even at very small timescales.

The heat flux $F_r$ at the radius $r$ is $F_r = -\kappa \partial T/\partial r$. The initial state is assumed to be isothermal with $T=10$ MeV. The boundary condition at the stellar center is $F_r\vert_{r=0}=0$ and the boundary condition at the stellar radius is $F_r\vert_{r=R}=(L_\text{Schwinger}+L_\text{Zakharov})/4\pi R^2$. The last condition means that total energy flux emitted from the surface has two terms: the first corresponds to the Schwinger mechanism and the second corresponds to the Zakharov mechanism. 

\begin{figure}[ht]
\includegraphics[width=\columnwidth]{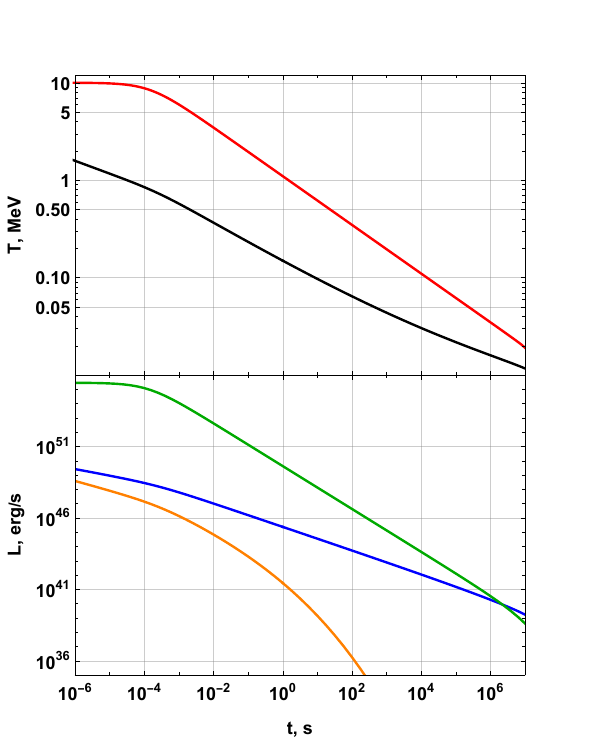}
\caption{Bottom: Neutrino luminosity (green), Zakharov luminosity (blue), and Schwinger luminosity (orange) as a function of time. Top: Core temperature (red), and surface temperature (black) as a function of time.}
\label{LTplot}
\end{figure}

The thermal evolution of the interior of the star depends on the phase of quark matter \cite{prakapenia2026a}. But it turns out that surface temperature evolves qualitatively in the same way for all phases: 2SC, 2SC+s and CFL (for large enough gap parameters $\Delta \simeq 50$ MeV), so in what follows we discuss just the 2SC+s case. The evolution of luminosity and temperature with time is shown in Figure \ref{LTplot}. The surface temperature evolves approximately as a power law $T\propto t^{-0.17}$.

Both photon and electron-positron pair luminosities fall exponentially at sufficiently long times. Indeed, for times $t\gg 1$ s, Schwinger luminosity can be neglected, and the surface thermal evolution is determined solely by bremsstrahlung emission. Photons dominate over neutrinos for $t>10^6$ s. At very large times the temperature gradient at the surface disappears, and neutrino emission again becomes dominant with respect to photon emission. 

The electrosphere itself is collisionless and transparent for photons. Nevertheless, above the temperature $T_{tr}\simeq 22$ keV, there are enough photons with energies exceeding the pair production threshold, to produce electrons and positrons (in addition to those produced by the Schwinger process), thereby making the medium opaque. Given very large opacity, photons and pairs thermalize \cite{2007PhRvL..99l5003A} at microphysical distance above the surface of a quark star with a temperature $T_0$ determined from the total luminosity of photons and pairs, roughly equal to the surface temperature.

This optically thick electron-positron-photon plasma expands outward with acceleration due to its radiative pressure, with the bulk Lorenz factor $\Gamma$ increasing linearly with distance from the surface $\Gamma\propto r$ and comoving temperature $T_c$ decreasing adiabatically $T_c\propto 1/r$ \cite{2012arXiv1205.3512R}. The density of electron-positron pairs in thermal equilibrium becomes exponentially small at large radii due to Boltzmann suppression. 

The estimate of the temperature when expanding plasma becomes transparent can be obtained as follows. The number density of photons above the threshold for pair production is $n_{\gamma,tr}=\pi^{-2}\int_1^\infty\frac{x^2}{\exp(x/\theta)-1}dx\simeq \pi^{-2}\theta\exp(-1/\theta),$
where $\theta=T/m_e$ is the dimensionless temperature. The cross section for $\gamma\gamma$ pair production (Breit-Wheeler process) peaks above the threshold energy $m_e$ with the maximum value $\sigma_{\gamma\gamma}\simeq \sigma_T/6$, where $\sigma_T$ is the Thomson cross section. The photon number density in the relativistic wind decreases as $r^{-3}$, so the optical depth is
$\tau=\int_R^\infty n_\gamma(r)\sigma_{\gamma\gamma} dr=\frac{1}{2}n_{\gamma,tr}\sigma_{\gamma\gamma}R.$
The condition $\tau=1$ leads to the following equation
\begin{equation}
    \frac{2}{9\pi}\alpha^2\theta\exp(-1/\theta)\frac{R}{m_e}=1,
    \label{OpT}
\end{equation}
which for $R=10$ km gives a solution $\theta_{tr}\simeq0.045$ or $T_{tr}\simeq 22$ keV. The same result may be obtained by considering Thomson scattering of photons on thermal electron-positron pairs \cite{2013ApJ...772...11R}.

When the comoving temperature decreases to the value $T_{tr}$ \cite{1998MNRAS.300.1158G} photons can again stream freely, with a spectrum very close to black body radiation \cite{2013ApJ...772...11R}. The corresponding radial distance from the surface of the star where this happens $R_{ph}\simeq R T_0/T_c\gg R$ is called the photospheric radius. Therefore, we expect that at early times quark star is characterized by black body electromagnetic radiation, generated near the electrosphere and emitted from its relativistic photosphere.

As the surface cools below $T\lesssim22$ keV ($t\gtrsim10^5$ s) there remain very few photons with energy exceeding the kinematic threshold for pair production. Then bremsstrahlung photons stream mostly freely outside of the electrosphere. Hence we expect that at sufficiently large times the quark star emits photons with the spectrum originating at the electrosphere. This spectrum produced by the Zakharov mechanism is similar to a part of the black body cut below the energy $\omega_{\min}\simeq 40$ keV. At these low temperatures, the spectrum is essentially reduced to a narrow peak at $\omega_{\min}$, see Figure \ref{Zspectra}. More quantitative information on transition from black body to the narrow peak can be obtained from kinetic simulations similar to the ones presented in \cite{2004ApJ...609..363A}, but this is beyond the scope of this Letter.

Finally, at very low temperatures $T\ll 22$ keV bremsstrahlung photons have so low energies that they are completely cut off by the plasma frequency of electrons in the electrosphere, and a newborn quark star becomes dark.

\textit{Discussion and conclusions. --} In summary, we revisited the thermal evolution and observational appearance of a newborn quark star. The surface cools down as a result of bremmstrahlung emission from the electrosphere, while the interior cools down due to neutrino emission. The photon luminosity decreases as a broken power law from $10^{50}$ erg/s at $10^{-7}$ s to $10^{49}$ erg/s at $10^{-5}$ s up to $10^{40}$ erg/s at $10^{6}$ s, and then exponentially.

If the birth of a quark star is directly observable, it is seen as a brief and bright flash of black body radiation originating from the relativistic photosphere at radii $R\gg 10^6$ cm with temperature decreasing below 1 MeV in a fraction of ms. The spectrum remains similar to a black body as the temperature decreases further to 22 keV at $t\simeq 10^5$ s. At even smaller temperatures, the spectrum transforms into a narrow peak, centered at $40$ keV. This result is in sharp contrast with previous studies \cite{2002PhRvL..89m1101P, 2004ApJ...609..363A} based on the assumption that the surface of a bare quark star cools due to the Schwinger pair creation.

This spectral feature is produced by a compact astrophysical object possessing an electrosphere, which is a characteristic feature of quark stars. We expect that such observatories as Swift BAT and Fermi GBM following the transient phenomena, as well as INTEGRAL and NuSTAR and planned XEUS observatories will be able to detect these objects.
Taking for the typical limiting flux for 1-day exposure as $F\sim 10^{-13}$ erg s$^{-1}$cm$^{-2}$ and using the luminosity distance formula $L=4\pi FD_L^2$ the limiting distance for detection of newborn bare quark star is about 100 Mpc.

{\bf Acknowledgements.}  This work is supported within the joint BRFFR-ICRANet-2025 funding programme under the grant No. F25ICR-001. We thank B.G. Zakharov for discussions.

\bibliography{total}{}
\bibliographystyle{h-physrev}

\clearpage

\end{document}